\documentclass[a4paper,10pt]{article}
\usepackage[latin1]{inputenc}
\usepackage[english]{babel}
\usepackage{latexsym}
\usepackage{graphicx}
\usepackage{multirow}
\usepackage{xspace}
\usepackage{fullpage}
\usepackage{amsmath}
\usepackage{amsfonts}
\usepackage{amssymb}
\usepackage{url}
\usepackage{authblk}
\usepackage[numbers,sort&compress]{natbib}
\newcommand{\bdel}{$\beta$-delayed\xspace}
\newcommand{\Hetub}{$^{3}He$\xspace}
\begin{document}
\title{\bf Conceptual design of the BRIKEN detector: \newline A hybrid neutron-gamma detection system for nuclear physics at the RIB facility of RIKEN}

\author[1,2]{A. Tarife\~no-Saldivia\footnote{Email: atarisal@gmail.com, ariel.esteban.tarifeno@upc.edu}}
\author[2]{J. L. Tain}
\author[2]{C. Domingo-Pardo}
\author[1]{F. Calvi\~no}
\author[1]{G. Cortes}
\author[3,4]{V. H. Phong}
\author[1]{A. Riego}
\author[ ]{ The BRIKEN collaboration\footnote{www.wiki.ed.ac.uk/display/BRIKEN/Home}}

\affil[1]{\emph{Universitat Polit\`ecnica de Catalunya (UPC), Barcelona, Spain}}
\affil[2]{\emph{Instituto de Fis\'ica Corpuscular (CSIC-University of Valencia), Valencia, Spain}}
\affil[3]{\emph{Department of Nuclear Physics, Faculty of Physics, VNU University of Science,Hanoi}}
\affil[4]{\emph{RI Physics Laboratory, RIKEN Nishina Center for Accelerator-based Science, Japan}}

\maketitle

\begin{abstract}
BRIKEN is a complex detection system to be installed at the RIB-facility of the RIKEN Nishina Center. It is aimed at the detection of heavy-ion implants, $\beta$-particles, $\gamma$-rays and $\beta$-delayed neutrons. The whole detection setup involves the Advanced Implantation Detection Array (AIDA), two HPGe Clover detectors and a large set of 166 counters of 3He embedded in a high-density polyethylene matrix. This article reports on a novel methodology developed for the conceptual design and optimisation of the 3He-tubes array, aiming at the best possible performance in terms of neutron detection. The algorithm is based on a geometric representation of two selected parameters of merit, namely, average neutron detection efficiency and efficiency flatness, as a function of a reduced number of geometric variables. The response of the detection system itself, for each configuration, is obtained from a systematic MC-simulation implemented realistically in \textsc{Geant4}. This approach has been found to be particularly useful. On the one hand, due to the different types and large number of 3He-tubes involved and, on the other hand, due to the additional constraints introduced by the ancillary detectors for charged particles and gamma-rays.
Empowered by the robustness of the algorithm, we have been able to design a versatile detection system, which can be easily re-arranged into a compact mode in order to maximize the neutron detection performance, at the cost of the gamma-ray sensitivity. In summary, we have designed a system which shows, for neutron energies up to 1(5)~MeV, a rather flat and high average efficiency of 68.6\%(64\%) and 75.7\%(71\%) for the hybrid and compact modes, respectively. The performance of the BRIKEN system has been also quantified realistically by means of MC-simulations made with different neutron energy distributions. 
\end{abstract}

\section{Introduction}\label{sec:intro}
Beta-delayed neutron emission is a process occurring when the neutron separation energy in the daughter nucleus ($S_n$) becomes smaller than the energy window for the beta-decay ($Q_{\beta}$). For neutron-rich nuclei far from the stability valley, beta-delayed neutron emission is the dominant decay process. The probability of a nucleus to emit a delayed neutron ($P_n$) yields information about the distribution of the nuclear levels populated in the decay \cite{borzov2005_beta}. It is worth mentioning that most of the $\sim 3000$ nuclei not discovered yet are expected to be $\beta$-delayed neutron emitters. In addition to the importance for nuclear structure, $P_n$ values are among the most important physical inputs for $r$-process model calculations \cite{arconesetal2011_dynamical}. Determining many yet unmeasured $P_n$ values is critical for testing the physical conditions of the astrophysical environment, and to provide clues about the site of the $r$-process. Consequently, the measurement of beta-delayed neutron emission probabilities provides unique information about the nuclear structure in exotic nuclei and contributes to the understanding of the formation of the heavy elements in the universe. Finally, $P_n$ values are also important for nuclear technology applications. Beta-delayed neutron emission plays an important role on the transient reponse of nuclear reactors and the quantification of decay heat \cite{NDC0683}. Therefore, future reactor designs using advanced fuel compositions will benefit from the improvements on current uncertainties of $P_n$ values.

The most common approach to measure $P_n$ values is by registering the $\beta$-decays of an specific nucleus in a $\beta$ detector ($N_{\beta}$) and the delayed neutrons in a neutron detector ($N_{n}$), respectively. Thus, for single delayed neutron emission one can write \cite{agramuntetal2015_characterization}
\begin{equation}
\label{eq:BetaNeutronMethod}
 P_n= \frac{\bar \varepsilon_{\beta}}{\bar \varepsilon_{n}} \frac{N_{n}}{N_{\beta}}.
\end{equation}
In this equation, $\bar \varepsilon_{\beta}$ and $\bar \varepsilon_{n}$ are the beta and neutron detection efficiencies respectively, averaged over all $\beta$ and neutron energies. 

Whenever the neutron background rate is comparable or higher than the delayed neutron rate, the quantity to be measured is the number of neutrons emitted simultaneously or in time coincidence with the $\beta$ particles ($N_{\beta n}$). In that case equation \ref{eq:BetaNeutronMethod} reads as, 
\begin{equation}
\label{eq:BetaNeutronCoincMethod}
P_n= \frac{\bar \varepsilon_{\beta}}{{\bar \varepsilon_{\beta}}' \bar \varepsilon_{n}} \frac{N_{\beta n}}{N_{\beta}},
\end{equation}
where ${\bar \varepsilon_{\beta}}'$ corresponds to the beta efficiency averaged over the neutron unbound states and $\bar \varepsilon_{\beta}$ is the beta efficiency averaged over all states.

Examination of equations \ref{eq:BetaNeutronMethod} and \ref{eq:BetaNeutronCoincMethod} shows that the measured $P_n$ values are strongly dependent on the response of the beta and neutron detectors employed. 
One of the most common methods to detect neutrons is based on the use of 3He-counters embedded in a polyethylene matrix~\cite{reederetal1977_average}. In this case, the response of the system has to be optimized in such a way that $\varepsilon_{n}$ becomes as less sensitive as possible to the primary neutron energy distribution. This condition needs to be fulfilled over a sufficiently broad neutron-energy range, comparable to the expected Q$_{\beta n}$-values involved in the $\beta$-decays under study. In addition, the average neutron detection efficiency needs to be maximized for the measurement of very exotic species, whose production yields are typically very low. These two requirements can be best accomplished if a sufficiently large number of 3He counters becomes available.

To this aim, a joint international collaboration has been stablished, which should enable an ambitious programme for Beta-delayed neutron measurements at RIKEN (BRIKEN) \cite{brikenCol}. The programme is focused on the first measurement of many single neutron emitters in the mass region from N$\sim$50 up to N$\sim$150, high-accuracy $P_n$ measurements of several known nuclei and first measurements of multiple neutron emitters (2n, 3n, \ldots). 

The radioactive isotopes of interest will be produced with the Rare Isotope Beam Factory (RIBF). At RIBF, $^{238}$U beams are accelerated by the Superconducting Ring Cyclotron (SRC) up to energies of $345 MeV/u$ and strike the production target at the entrance of the BigRIPS fragment separator. The first stage of BigRIPS is employed for selection and purification of the primary beam by means of the $B\rho - \Delta E - B\rho$ method. In the second stage, the fragments are identified by measuring their magnetic rigidity $B\rho$, their energy loss $\Delta E$  inside an ionization chamber and their time-of-flight (TOF) by using plastic scintillators. The secondary beam is then slowed down by means of an aluminum degrader, focalized and stopped in a stack of six double-sided silicon-strip detectors called AIDA (Advanced Implantation Detector Array)\cite{AIDA}. The latter is axially surrounded by the BRIKEN neutron detector array, which consists of more than 160 3He tubes, of up to six different types, embedded in a large PE-block, as it is described below. Two auxiliary HPGe clover detectors can be optionally inserted from both sides in order to measure also $\gamma$-rays following the beta-decay. Thus, it is the world largest detector of its kind for the detection of beta delayed neutrons. The BRIKEN detector builds upon the experience gained from previous experiments, such as BELEN at JYFL (Finland) \cite{agramuntetal2014_New,agramuntetal2015_characterization} and GSI (Germany) \cite{caballero-folchetal2014_beta}, 3Hen at ORNL (USA) \cite{grzywaczetal2014_hybrid} and NERO at NSCL-MSU (USA) \cite{pereiraetal2010_neutron}.

In order to cope with the large variety of constituents, the compactness and the geometrical complexity of the full detection ensemble, a robust algorithm has been specially developed in order to design and optimize the distribution of \Hetub counters inside the HDPE-matrix. This article describes a new topological method, which is based on a geometric representation of two selected parameters of merit, namely, average neutron detection efficiency and efficiency flatness, as a function of a reduced number of geometric degrees of freedom. 

This present work is organized as follows. The main performance requirements and hardware available for BRIKEN are summarized in Sec.~\ref{sec:description}. 
In Sec.~\ref{sec:design} are described the topological algorithm, simulation code, figures of merit used within this work,  together with symmetry aspects related to neutron transport in the PE moderator. Sec.~\ref{sec:hybridDesign} presents the optimization study for the hybrid setup. The impact of the neutron-energy distribution is quantified in Sec.~\ref{sec:impact} by means of MC simulations corresponding to extreme cases. A second iteration of the optimizacion for design of the compact mode is presented in Sec.~\ref{sec:compact}. The final configuration for the BRIKEN neutron detector is discussed in Sec.~\ref{sec:finalConf}. Finally, a discussion of the BRIKEN neutron detector with respect to other setups and the main results and conclusions of this work are presented in 
Sec.~\ref{sec:discussion} and ~\ref{sec:summary}, respectively.

\section{Main performance requirements and available materials}\label{sec:description}

In order to avoid that the overall detection sensitivity of the full system is limited by the performance of the neutron detector itself, it is necessary that the neutron detection efficiency becomes, at least, comparable to the beta-detection efficiency. In this respect, an average efficiency of 60\% is defined as the minimal target value for the present design study.

For nuclei with large Q$_{\beta n}$ values populating states at high-excitation energies in the daughter nucleus, a strong dependency of the detector efficiency with the neutron energy may become important for the measurement of the $P_n$ value  (see eqs. \ref{eq:BetaNeutronMethod} and \ref{eq:BetaNeutronCoincMethod}). However, these kind of transitions are readily suppressed by the effect of the Fermi function. Therefore, accordingly to the previous experience \cite{agramuntetal2015_characterization}, a nearly constant efficiency up to neutron energies of 1~MeV, and small variations up to 5~MeV, represents a reasonable assumption. This premise will be confronted with extreme hypothesis for neutron energy spectra (see Sec.~\ref{sec:impact}) in order to quantify its possible impact on the performance of the final BRIKEN detection system.

In addition, for the physics programme of BRIKEN it has been found desirable to have the possibility of gamma-ray detection, whenever this is compatible with high neutron efficiency and flatness (hybrid mode).

In summary, the main performance requirements for the BRIKEN neutron detector are:
\begin{enumerate}
 \item Neutron efficiency higher than 60\% up to 1~MeV.
 \item Flat response up to 1~MeV and small variations of the efficiency up to 5~MeV.
 \item Gamma-ray detection capabilities compatible with high neutron efficiency and flatness (hybrid mode).
\end{enumerate}

To fulfill the requirements of high efficiency and flat response, a large amount of \Hetub-filled neutron counters are mandatory. \Hetub-filled tubes are very efficient counters for thermal neutrons because of the high cross section for the capture reaction $^{3}He(n,p){}^{3}H$ ($\sigma_{th}=5345~barn$, $Q=764~keV$) \cite{chadwicketal2011_ENDF}. At higher energies, neutron moderation is necessary to obtain high detection efficiencies. The properties of the \Hetub  tubes available in the BRIKEN collaboration are listed in table \ref{tab:TAB_3He_briken_coll}. Some of these tubes have been used in previous setups for \bdel neutron detectors such us BELEN, 3Hen and TETRA. 

A moderator of High-Density Polyethylene (HDPE) is used for the BRIKEN detector. The moderator has a total size of $90\times 90 \times 75~ cm^3$ and is composed by 15 HDPE slices of 5~cm thickness. The slices are assembled together by using stainless steel rods passing along the corners of each slice. Depending on the experimental background conditions, the moderator will be shielded using additional HDPE slices and cadmium layers.

Gamma-ray detection capabilities are implemented at BRIKEN  by two large volume germanium detectors (clovers) embedded into the HDPE moderator. This mode is known as \textbf{hybrid mode}. The clover detectors are from the CARDS array type and allow for high precision gamma spectroscopy. Each clover has a total photopeak efficiency $\sim 1\%$ at 1.33MeV \cite{krolasetal2002_first}. Because the neutron efficiency may be limited by the clovers, flexibility to transform the hybrid mode into a $4\pi$ neutron counter is also a design requirement of the BRIKEN detector. This flexibility is achieved by a mechanical setup allowing to remove  clover detectors easily in order to fill up the empty space with HDPE plugs and additional \Hetub-tubes. This operational mode is known as \textbf{compact mode}.

A special Data Acquisition system (DACQ) has been developed for the BRIKEN neutron counter. A detailed description of this system is provided in references \cite{brikenDAQ,agramuntetal2015_characterization}. The DACQ is a self-triggered data acquisition system based on VME digitizers type SIS3302 (8 channels, 100 MSamples/s, 16bit) and SIS3316 (16 channels, 250 MSamples/s, 14bit); both from Struck Innovative Systeme \cite{struck}. The DACQ is able to handle with multiple VME crates and up to 192 channels running independently. By using the on-board data processing of the digitizers, a timestamped readout with the signal amplitude is obtained at each channel. The timestamp is provided by a common time reference distributed to all involved detection systems (BRIKEN counter, AIDA and Beam detectors). 

\begin{table}[t]
\setlength{\tabcolsep}{3pt}
\let\center\empty
\let\endcenter\relax
\centering
\caption{Description of available \Hetub -filled tubes for the BRIKEN neutron counter.}
\label{tab:TAB_3He_briken_coll}
\resizebox{\width}{!}{\newlength{\hcolw}
\setlength{\hcolw}{40pt}
\begin{tabular}{ccccccccc}
\noalign{\smallskip}\hline\noalign{\smallskip}						
\multirow{3}{\hcolw}{\centering  Detector type} & \multirow{3}{*}{\hfil Manufacturer}	& \multirow{2}{*}{\centering  Pressure}		& \multirow{2}{\hcolw}{\centering  Tube diameter}	& \multirow{2}{\hcolw}{\centering   Tube length}	& \multirow{2}{\hcolw}{\centering   Active length}	&\multirow{3}{\hcolw}{\centering Previous setup} 	&\multirow{2}{\hcolw}{\centering  Number of counters}	\\
&						&							&							&							&							& 								& 								\\
&						& (atm)							& (inch/cm)						& (inch/cm)						& (inch/cm)						& 								& 								\\
\noalign{\smallskip}\hline\noalign{\smallskip}						
\bf B &\multirow{2}{2.0cm}{LND Inc.}			& 8							& 1.0/2.54						& 26.61/67.6						& 23.62/60.0						& \multirow{2}{\hcolw}{\centering BELEN \cite{agramuntetal2014_New,agramuntetal2015_characterization,gomez-hornillosetal2014_beta}}								& 42								\\
\bf R &						& 10							& 1.0/2.54						& 26.61/67.6						& 23.62/60.0						& 								& 10								\\
\cline{1-7}					
\bf I &\multirow{3}{2.0cm}{GE  Reuter Stokes}		& 10							& 1.0/2.55						& 29.37/74.6						& 24.0/60.96						&\multirow{2}{\hcolw}{\centering 3Hen \cite{grzywaczetal2014_hybrid}} 								& 17								\\
\bf K &						& 10							& 2.0/5.08						& 27.48/69.79						& 24.0/60.97						& 								& 64								\\
\bf E &						& 5							& 1.0/2.54						& 15.6/39.63						& 11.18/30.0						& ---								& 26								\\
\cline{1-7}						
\bf N &---						& 4							& 1.18/3.0						& 23.62/60						& 19.69/50.0						& TETRA \cite{testov2015response}								& 20								\\
\cline{1-7}						
\multicolumn{6}{c}{Total counters for the BRIKEN detector}																										& 								& 179							\\
\noalign{\smallskip}\hline\noalign{\smallskip}						
\end{tabular}}
\end{table}

\section{The topological Monte Carlo algorithm}\label{sec:design}

Apart from the large number of detection elements, the precise geometric arrangement of the 3He-tubes represents the key feature to achieve both high and flat neutron detection efficiency over a broad neutron energy range. The optimization of a system combining different tube sizes and diameters, and such a large number of counters is a rather complex and time consuming task. Extensive MC simulations are required and the optimized solution is not necessarily the unique one. In addition, technical constraints such as construction tolerances, minimum distance between tubes for sufficient mechanical stability and tubes alignment make the optimization problem even more complex and challenging. 

In order to deal with these difficulties, a methodology based on the optimization of a geometrically parameterized counter array distribution is proposed. In first instance, the parameters of merit are defined (see Sec.~\ref{sec:parMerit}) in order to guide the selection of the most convenient configuration, according to some specific physics requirements. Additionally, the shape of the array builds upon the symmetries associated to the process of transport and moderation of neutrons in the HDPE (see Sec.~\ref{sec:emptyMod}). Secondly, a mathematical effort is carried out (see Sec.~\ref{sec:Par_hybrid}) in order to obtain an analytical geometric description, which describes any feasible distribution of a certain number of 3He-tubes within a given HDPE-volume. For this to become effective, the geometry model needs to have as less parameters as possible, typically between two and three. 

The Monte Carlo simulations needed for this work have been implemented in \textsc{Geant4} version 10.0.3 \cite{agostinellietal2003}. Data libraries provided by default with this version have been used. The calculations were carried out on a desktop computer running on UBUNTU 14.04 with g++ version 4.8.4. Currently, \textsc{Geant4} calculations using neutron proportional counters moderated with Polyethylene results in good agreement with MCNPX calculations and experimental measurements. In past versions, there was a systematic bias in the calculated efficiency due to bugs in the interpolation routines of the thermal libraries \cite{garciaetal2013}. Final corrections to these bugs have been introduced in versions 10.0 and later. Benchmark calculations between  \textsc{Geant4} (10.0.3) and MCNPX (version 2.5) for a simplified BRIKEN geometry have yielded relative differences in the efficiency of less than 1\% \cite{tarifeno-Saldiviaetal2015_benchmark}.

Because the use of two large volume HPGe clover detectors, the hybrid configuration is the most complex case in this study. Therefore, we start the design study with the hybrid configuration. Counters of type B to E have been used for this set-up. Once a suitable hybrid set-up has been found, in a second iteration, the compact configuration design will be tackled.  In this case, tubes of type N have been added to the optimal configuration for the hybrid mode; a further optimization of tubes type N in the counter array has been then carried out in order to obtain an optimal configuration for the compact mode. 

\subsection{\textsc{Geant4} detector model and simulations}

\begin{figure}[t]
\includegraphics[width=\textwidth,keepaspectratio]{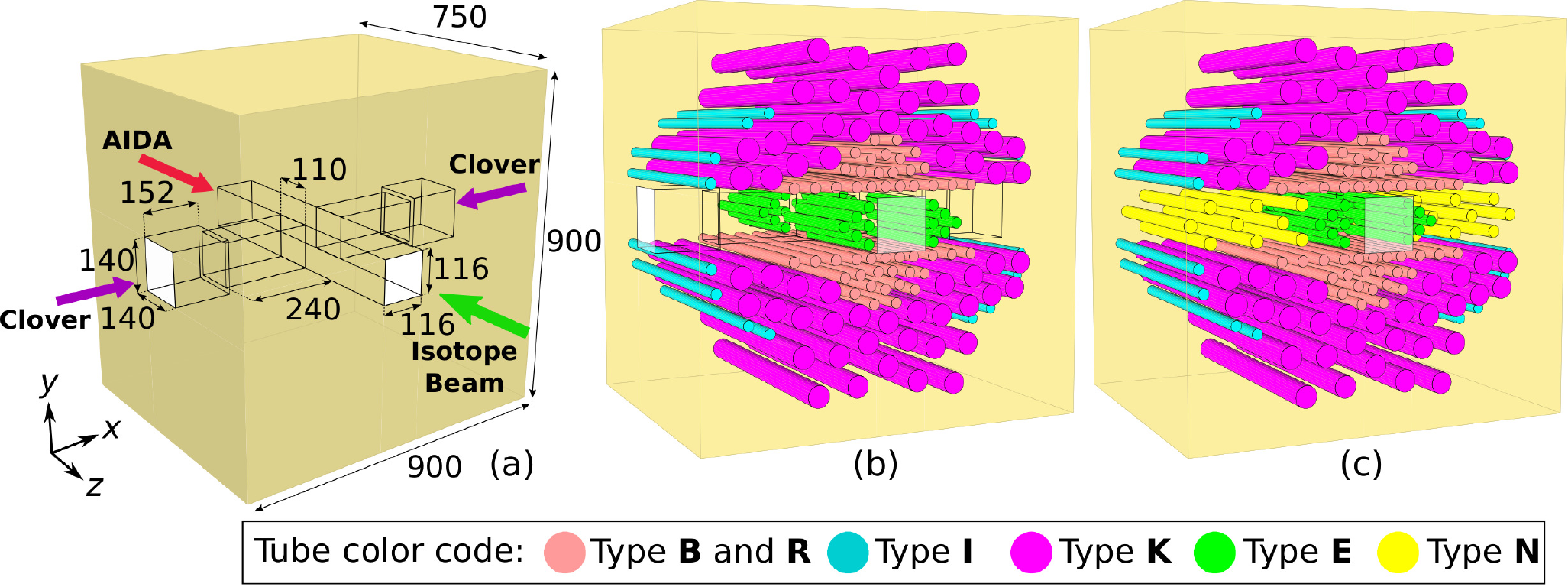}
\caption{\label{fig:G4detGeometry} Detector geometries implemented in  \textsc{Geant4}. (a) HDPE moderator geometry for the hybrid mode, dimensions in $mm$. Clover detectors are embedded at each side of the moderator along the X-axis. AIDA detector is embedded along the Z-axis. (b) and (c) correspond to the counter array distributions for the hybrid (148 tubes) and compact (166 tubes) modes, respectively. Because of visualization purposes, only active volumes are displayed in (b) and (c).}
\end{figure}

A generic geometry implemented in \textsc{Geant4} is presented in figure \ref{fig:G4detGeometry}. In particular Fig.1.a shows the moderator block itself for the hybrid set-up, where the internal squared holes are designed to host both the clover HPGe detectors and AIDA. Fig.1.b and Fig.1.c show generic representations of the counter array distributions for the hybrid and compact geometries, respectively. In the detector model, the HDPE density is $0.95~g/cm^3$ and the room temperature is set to $293.6~K$. The simulation of \Hetub counters is implemented using active and passive volumes \cite{tarifenoetal2013_modelling}. The material and dimensions for each tube type is defined according to the information provided by the manufacturer in the data sheet. In the simulation, detected events correspond to energy deposition from $150~keV$ up to $900~keV$ inside the \Hetub active volume.
A particle generator for mono-energetic neutrons has been implemented. The generator is also able to reproduce the energy distributions taken from data libraries or experimental data. The neutrons are emitted isotropically from a point source placed at the center of the moderator (see fig. \ref{fig:G4detGeometry}). For the optimization of the counter array, calculations have been carried out for a set of discrete neutron energies, namelly
 \begin{equation}
 \label{eq:Energies}
  E_i (MeV)=\{0.001, 0.001, 0.01, 0.1, 1.0, 2.0, 3.0, 4.0, 5.0\}.
 \end{equation}

\subsection{Parameters of merit}\label{sec:parMerit}
The neutron efficiency is calculated from the ratio $\eta_i(E_i)=$(Detected events)/(Processed events). Two figures of merit are defined as a function a maximum energy limit ($E_{max}$): 
 \begin{itemize}
 \item{Average efficiency:} Correspond to the average for the set of discrete energies used in the calculations, 
 \begin{equation}
 \label{eq:averageEff}
  \big \langle \eta \big \rangle (E_{max}) = \sum_{E_i \leq E_{max}} \frac{\eta(E_i)}{\#\{E_i\}}
 \end{equation}
 \item{Flatness factor:} A measure of the flat response until a maximum energy limit ($E_{max}$),
 \begin{equation}
  \label{eq:flatness}
  F(E_{max}) = \frac {Max(\eta(E_i))}{Min(\eta(E_i))}, E_i \leq E_{max}
 \end{equation}
\end{itemize}

A counting statistics is assumed for estimation of uncertainties in the efficiency. Thus, uncertainties associated to the flatness are estimated from error propagation for the quotient.

\subsection{Symmetry aspects related to neutron transport}
\label{sec:emptyMod}

\begin{figure}[t]
\begin{center}
\includegraphics[width=\textwidth,keepaspectratio]{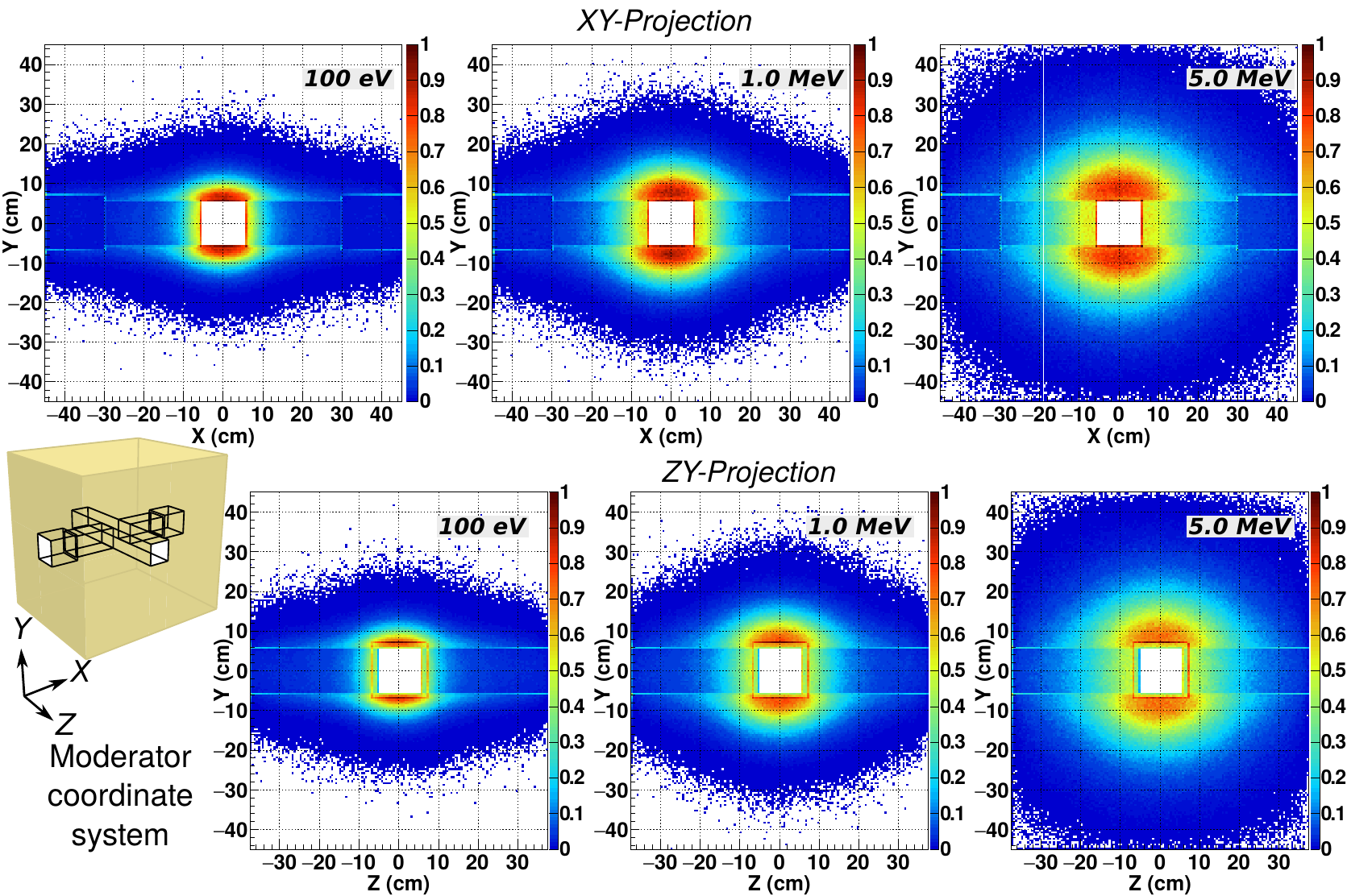}
\caption{\label{fig:emptyModPlots} 
Normalized neutron density inside the HDPE for moderated energies $E_n \leq 1eV$. Initial neutron energy from left to right: $100 eV$, $1.0MeV$, $5.0 MeV$. The neutron density integrated onto the $Z$ and $X$-\emph{axis} ($XY$ and $ZY$-\emph{projection}) is shown on top and bottom, respectively. }
\end{center}
\end{figure}

Neutrons in the moderator undergo a reduction of their initial energy by multiple scattering with hydrogen nuclei. As a rule of thumb, the capture reactions inside the \Hetub-counters starts to be significant for moderated energies below $\sim 1 eV$. The neutrons can be also captured  in the moderator through $(n,\gamma)$ reactions with hydrogen nuclei. Capture reactions in the wall of the counters are negligible provided the small cross sections of the constituent materials (aluminum and stainless steel). Some neutrons scape from the moderator without any capture reaction. As it can be appreciated in Fig.~2, the main losses of neutrons take place through the holes and 
the adjacent region for HPGe clover detectors and AIDA, respectively. Thus, the internal geometry of the moderator and the position of the counters in these region become relevant to the optimization of the BRIKEN detector performance.

In previous compact setups such as NERO, BELEN or 3HEN, the moderator has a central hole to accommodate an implantation system and a beta detector. For such geometry, the counters have been arranged in concentric rings around the central hole. This design concept is motivated by the cylindrical symmetry around the axis hole arising from the moderation of neutrons.

In the BRIKEN moderator, the symmetry around the central hole is modified by the clovers holes. In order to have insights on how to arrange the \Hetub-counters, a study of the transport and slowing down of neutrons in the BRIKEN moderator has been carried out. The tracking capabilities of the  \textsc{Geant4} toolkit are well suited for this purpose. For the study, $4 \times 10^6$ neutrons have been processed for the initial energies $100~eV$, $1.0~MeV$ and $5.0~MeV$. The neutrons have been tracked until they scape or are captured in the moderator. The final positions have been used to calculate the neutron density for moderated energies less than $1~eV$. The results of the study are presented in figure \ref{fig:emptyModPlots}. The neutron density has been integrated along the $Z$ and $X$-axis ($XY$ and $ZY$-projection respectively), and normalized to its maximum value. The effect on the symmetries of the moderation process of the clover holes and AIDA hole can be observed in figure \ref{fig:emptyModPlots} on the ($XY$-projection) and ($ZY$-projection), respectively.

In figure \ref{fig:emptyModPlots}, there are two regions where the neutron density is symmetric: the regions on top and bottom of the holes along the $Y$-axis, and the zone adjacent to the holes in middle of the moderator. On top and bottom regions, the neutron density has an oval shape. At $100~eV$, the neutron density is peaked up to $\sim 4cm$ from the border of the holes. In contrast at $5~MeV$, the peak of the density is broadened up to $\sim 15cm$ from the border. Beyond this region, the density decreases roughly to a constant value in the external part of the moderator. Consequently, while the counters placed close to the holes are critical to obtain a high efficiency, the counters distributed on the external parts of the moderator are important for a flat response up to $5~MeV$. It is worth mentioning that, for the first group of counters placed in the internal region of the moderator, the separation distance from the border of AIDA and clover holes will also play an important role. This is a consequence of the high density values in this region.

\subsection{Parameterization of the counter array distribution}\label{sec:Par_hybrid}

\begin{figure}[t]
\begin{center}
\includegraphics[width=0.85\textwidth,keepaspectratio]{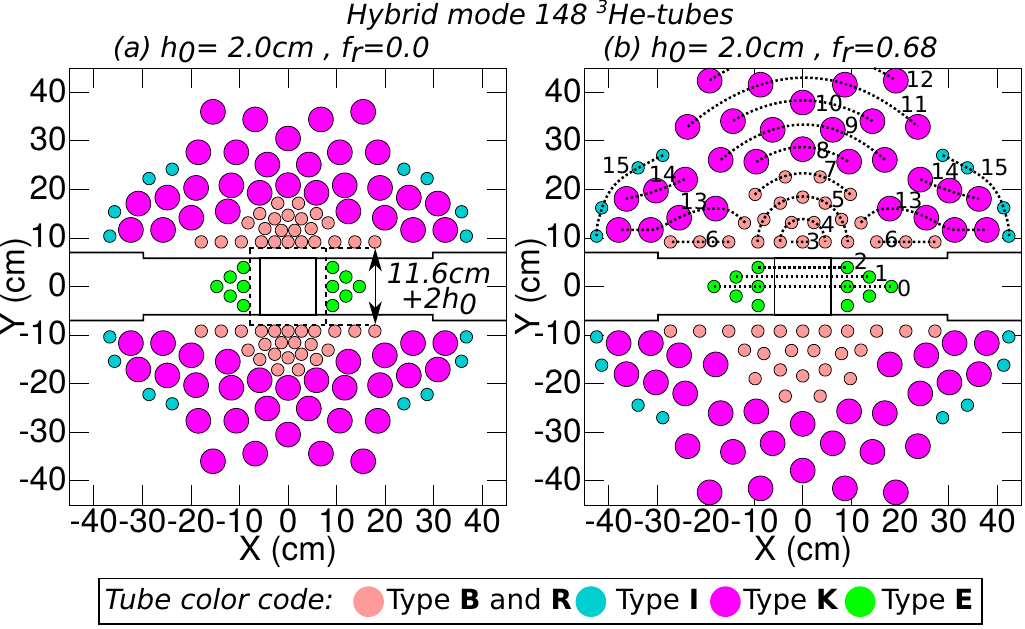}
\caption{\label{fig:parametrizationhybrid} Parameterized counter array distribution for the hybrid mode using 148 \Hetub-tubes. The size of the central hole is 11.6cm. The counter array distribution is determined by a set of parametric functions ($\vec R_i$) which depend on parameters $h_0$ and $f_r$. The parameterization is shown for parameters (a) $h_0=2cm$, $f_r=0.0$; and (b) $h_0=2cm$, $f_r=0.68$. The parametric functions  are depicted in (b) for the shell index $i:0 \ldots 15$ using dotted lines.
}
\end{center}
\end{figure}

The experience in previous setups has shown the importance of having tubes for spare. It becomes specially important during the experimental campaign, when beamtime cannot be wasted trying to find out how to solve problems in defective tubes (excessive noise, anomalous signals, etc.). Therefore, for the BRIKEN neutron detector around 5 to 10\% of the tubes by type have been reserved for spare. A total of 148 \Hetub-tubes of type B to E have been used in the counter array distribution for the hybrid mode. Including tubes type N, a total of 166 tubes have been used for the counter array in compact mode. 

The counter array distributions for the hybrid and compact mode are shown in figures \ref{fig:G4detGeometry}.b and \ref{fig:G4detGeometry}.c respectively. For the purpose of visualization, only the active volumes are visible in these  \textsc{Geant4} representations of the detector geometry. By symmetry arguments (see section \ref{sec:emptyMod}), several adoptions have been done for the parameterization of the counter array distribution:
\begin{enumerate}
 \item[i.]  All counters are placed along the $Z$-axis, i.e. in parallel with the AIDA hole axis.
 \item[ii.] The counter distribution has an oval shape.
 \item[iii.] The counter distribution is close-packed around the AIDA hole. Consequently, while small diameter counters are placed in the inner region of the moderator, higher diameter counters are placed in the external region.
\end{enumerate}
For the hybrid mode, as shown in figure \ref{fig:G4detGeometry}.b, tubes type B to K (see table \ref{tab:TAB_3He_briken_coll}) are placed on the top and bottom regions of the moderator. The shorter tubes type E are placed in the zone adjacent to the holes in middle of the moderator. These tubes are placed back and forward to the clover holes. 

For the compact mode, the clover holes are filled-up with PE. Collinear tubes type E are put in contact at the central plane of the moderator. In addition, tubes type N are placed next to tubes type E in the middle region, as shown in figure \ref{fig:G4detGeometry}.c. Tubes on top and bottom of the moderator remain in the same position as for the hybrid mode.

To facilitate the assembling of the detector in hybrid mode, the tubes are aligned in $Z$-axis at 5~cm of the moderator border. One of the 5 cm PE slices is used as the alignment point for this purpose. The only exception is for tubes type E, in that case border of the clover hole is used as the alignment point.

Motivated by mechanical reasons, a tolerance of 2~mm  has been assumed for the cavities housing the tubes in the moderator. In addition, a minimum distance of 5~mm between each cavity has been assumed to constrain the optimization of the counter array distribution. 

The parametrization for the hybrid mode using 148 \Hetub-tubes is presented in figure \ref{fig:parametrizationhybrid}. This counter array distribution has been achieved by the set of tubes ``shells'' depicted in figure \ref{fig:parametrizationhybrid}.b using dotted lines. The position of the shells in the $XY$-plane is determined by a set of parametric functions, 
\begin{equation}
 \vec R_i = \big( {F_i}_x (n_i, h_0, f_r),{F_i}_y (n_i, h_0, f_r)\big) , \qquad i: 0 \ldots 15.
\end{equation}
Where $n_i$ is the number of tubes in the shell, $h_0$ is the separation in \emph{cm} from the inner moderator border to the first shell next to the hole (see \ref{fig:parametrizationhybrid}.b); and $f_r$ is a dimensionless parameter which determines the spacing between adjoined shells. For a given shell, the tube spacing depends on $n_i$ and $f_r$. The effect of the variation of parameter $f_r$ on the shell separation and the tube spacing is observed by comparison of figures \ref{fig:parametrizationhybrid}.a and \ref{fig:parametrizationhybrid}.b.

\section{Hybrid BRIKEN designs}\label{sec:hybridDesign}
\begin{figure}[ht]
\begin{center}
\includegraphics[width=\textwidth,keepaspectratio]{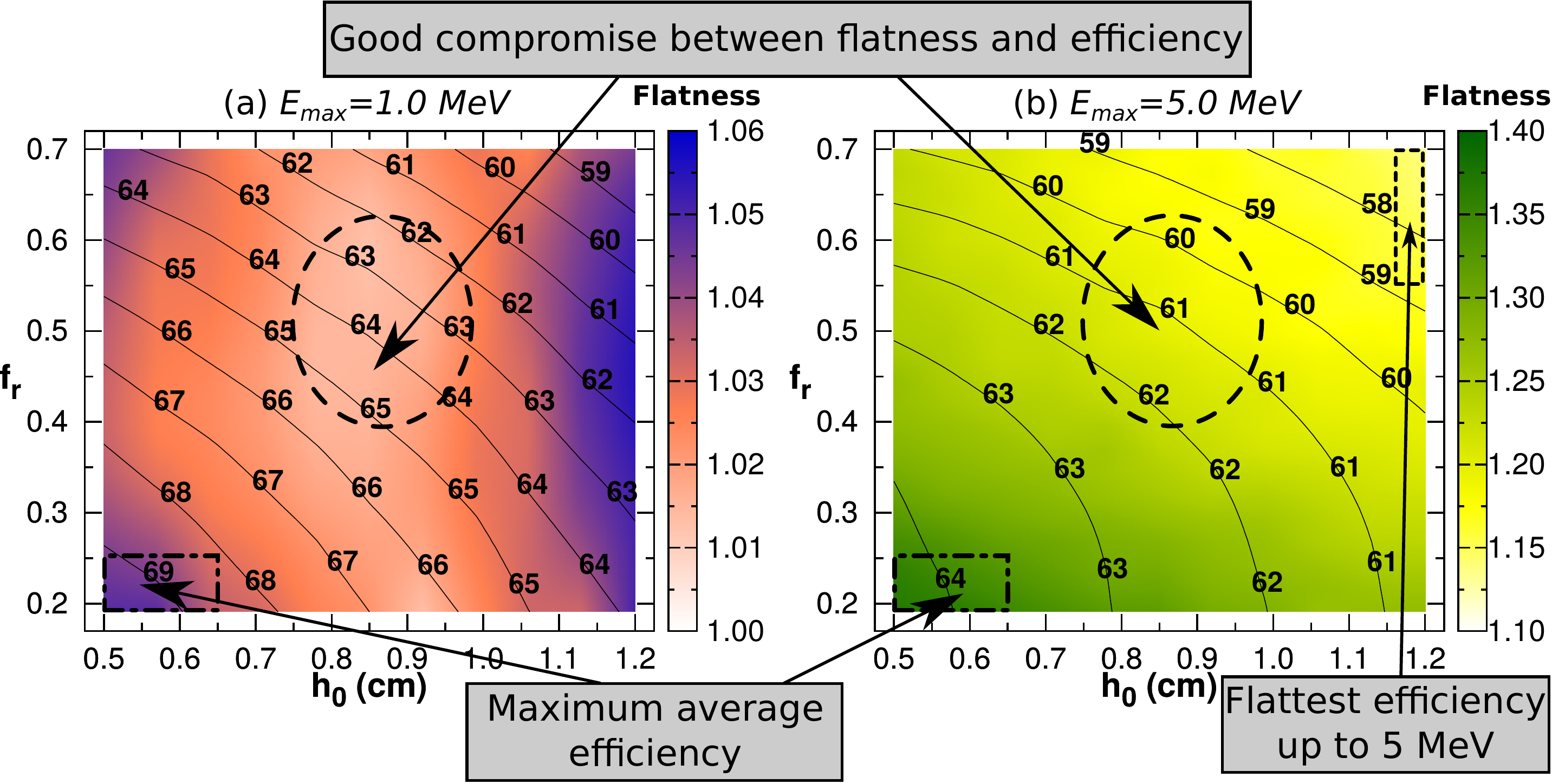}
\caption{\label{fig:ParSpace_hybrid} Average efficiency (eq. \ref{eq:averageEff}) and flatness (eq. \ref{eq:flatness}) in the parameter space $(h_0,f_r)$ for the hybrid mode. The average efficiency is displayed in percentage points with lines and the flatness as a colour map. In (a) results for $\big \langle \eta \big \rangle (1~MeV)$ and $F(1~MeV)$; in  (b) results for $\big \langle \eta \big \rangle (5~MeV)$ and $F(5~MeV)$. Regions of interest for the optimizacion in the parameter space are shown enclosed on both figures.}
\end{center}
\end{figure}
\begin{figure}[h]
\begin{center}
\includegraphics[width=\textwidth,keepaspectratio]{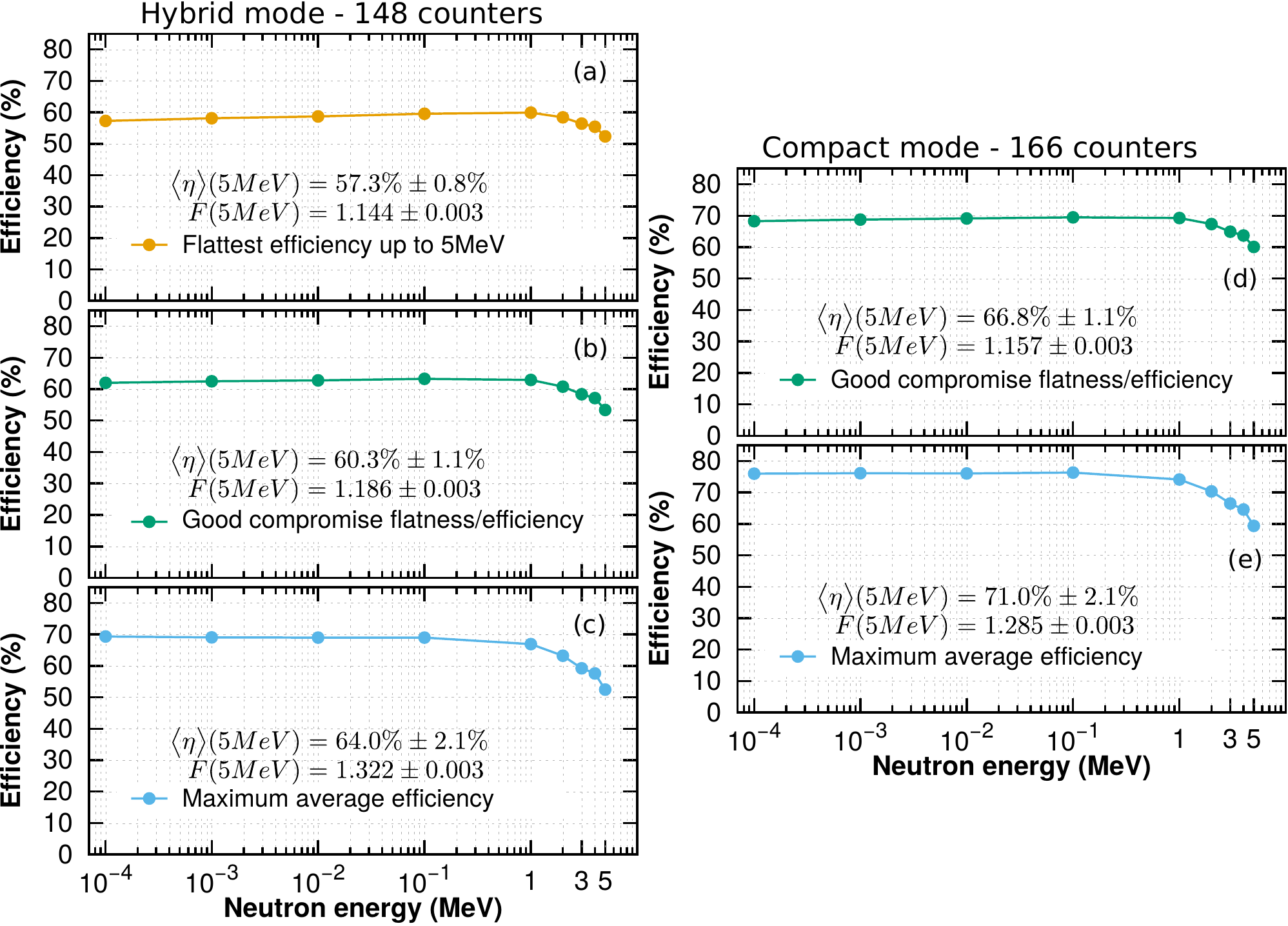}
\caption{\label{fig:NeutronSensitivityOptConf} Neutron responses for optimized configurations. (a), (b) and (c) hybrid mode using 148 counters. (d) and (e) compact mode using 166 counters.}
\end{center}
\end{figure}

Calculations of efficiency and flatness for the parametric counter array presented in section \ref{sec:Par_hybrid} have been carried out. The calculations are done on the parameter space $(h_0,f_r)$. Constraining the computations by the minimum separation between the tube housing cavities results in calculation ranges of $0.19 \leq f_r \leq 0.7$ and $0.5 ~cm \leq h_0 \leq 1.2 ~cm$. The  parameter space has been divided into a grid of $11\times 21$ points. A total statistic of $5\times10^4$ neutrons have been processed for each point in the grid. Such statistics allow one to identify regions of interest in the parameter space, while keeping the total computational time within a reasonable limit.  

The results of the calculations are presented in fig. \ref{fig:ParSpace_hybrid}. In these calculations, the typical statistical uncertainty for the flatness values are $\Delta F(1~MeV) \approx \pm 0.008$ and $\Delta F(5~MeV) \approx \pm 0.01$. For neutrons up to $1~MeV$, a very flat response ($F(1~MeV)< 1.1$) is obtained with the average efficiency ranging from  58\% up to 69\%. For neutrons up to $5~MeV$, a compromise between the average efficiency and flatness is observed, the greater the flatness the lower the efficiency. The average efficiency ranges from 56\% to 64\%, while the flatness value is $1.13 < F(5~MeV)<1.4$. Comparing calculations results up to $1$ and $5~MeV$, three regions of interest for optimization have been identified;  namely, (i) the region of flattest efficiency up to $5~MeV$, (ii) the region with a good compromise between efficiency and flatness up to $1$ and $5~MeV$, and (iii) the region of maximum efficiency up to $1$ and $5~MeV$. These regions are depicted with dashed lines in fig. \ref{fig:ParSpace_hybrid}. 

To proceed with the optimization, calculations with increased statistics ($5\times10^5$ processed neutrons) have been carried out for the configurations in the above mentioned regions of interest. By doing so the flatness uncertainty decreased to a level, which was low enough for the optimization purpose ($\Delta F(5~MeV) \lesssim \pm 0.003$). Finally, the optimized configuration for each region was obtained by maximizing the average efficiency and minimizing the flatness in the calculations up to $5~MeV$. The neutron responses for the three optimized configurations are shown in fig. \ref{fig:NeutronSensitivityOptConf}.

\section{Impact of the neutron energy distribution}\label{sec:impact}
\begin{figure}[t]
\begin{center}
\includegraphics[width=0.5\textwidth,keepaspectratio]{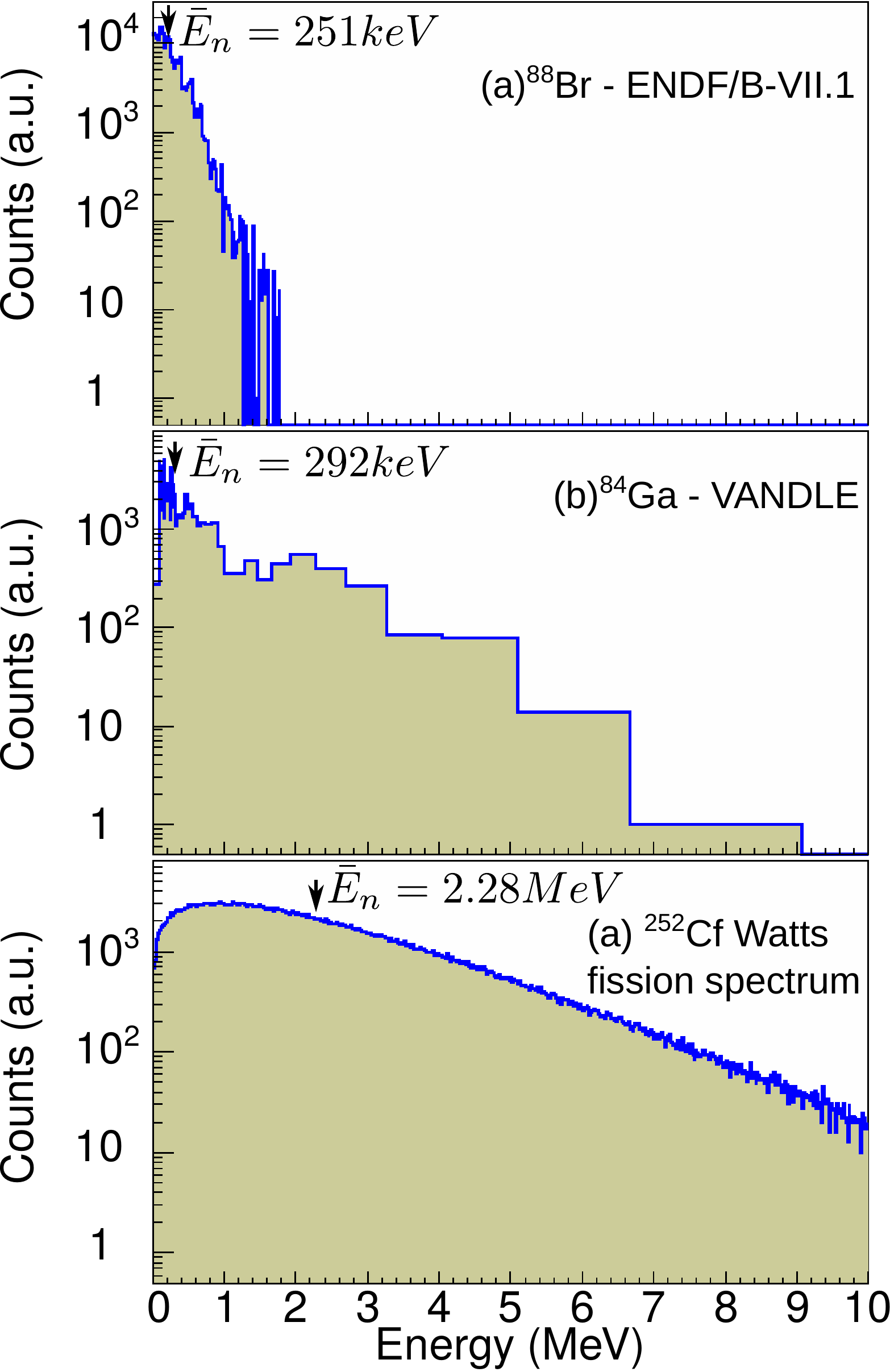}
\caption{\label{fig:nSpectra}Different neutron energy spectra simulated. (a) and (b) are representative of \emph{soft} and \emph{hard} beta-delayed neutron spectra. (c) ${}^{252}Cf$ neutron source from Watts~\cite{froehner1988_watt} fission spectrum.}
\end{center}
\end{figure}

\begin{table}
\begin{center}
 \caption{Response of the optimized hybrid configurations  to different neutron spectra.}
\label{tab:TABnSpectra}
\begin{tabular}{lccccc}
\noalign{\smallskip}\hline\noalign{\smallskip}
Configuration & \emph{F(5~MeV)} & $\dfrac{\eta({}^{84}Ga)}{\eta({}^{88}Br)}$ & $\dfrac{\eta({}^{252}Cf)}{\eta({}^{88}Br)}$ &  $ \eta_{1n}$ & $\eta_{2n}$\\ 
\noalign{\smallskip}\hline\noalign{\smallskip}
\emph{Flattest efficiency up to $5MeV$}& 1.144 & $1.000 \pm 0.003$ & $0.963 \pm 0.003$ &  $59.8\%$ & $35.8\%$\\ 
\emph{Good compromise flatness/efficiency}& 1.186 & $1.002 \pm 0.003$ & $0.947 \pm 0.002$ &  $63.3\%$ & $40.1\%$\\ 
\emph{Maximum efficiency up to $1$ and $5MeV$}& 1.322 & $0.996\pm 0.003$ & $0.908 \pm 0.002$ &  $68.6\%$& $47.0\%$\\ 
\noalign{\smallskip}\hline\noalign{\smallskip}
\end{tabular}
\end{center}
\end{table}

The response of the optimized configurations to different neutron spectra has been studied. This study aims to test the optimized configurations for representative neutron spectra and to identify which configuration is better suited for optimization of the compact mode. With this in mind, the efficiency has been calculated for two representative decays with a \emph{soft} and \emph{hard} neutron spectrum for medium-heavy nuclei. The \emph{soft} spectrum is shown in fig. \ref{fig:nSpectra}.a and corresponds to the ENDF/B-VII.1 spectrum of ${}^{88}Br$. The \emph{hard} spectrum is shown in fig. \ref{fig:nSpectra}.b and resembles the one measured for the decay of $^{84}$Ga with the VANDLE detector\cite{84GaVANDLE}. It is important to emphasize that both spectra have a similar average energy ($\bar E_n \approx 250-300~keV $), although a significant portion of the \emph{hard} spectrum extends beyond $1.5~MeV$. 

The efficiency for a ${}^{252}Cf$ neutron source (fig. \ref{fig:nSpectra}.c) has been also calculated. This spectra is too hard for a \bdel neutron emitter ($\bar E_n=2.28~MeV$), but it is commonly used for calibration of neutron detectors based on modetared proportional counters. Thus, the calculation of the ${}^{252}Cf$ efficiency provides a conservative estimation of the maximal deviation due to the hardness of the spectrum. In addition, it provides a reference of the expected efficiency from an experimental characterization of the detector. 

The results of the calculations are reported in table \ref{tab:TABnSpectra}. For each optimized configuration, the ratio of efficiencies $\eta({}^{84}Ga) / \eta({}^{88}Br)$ (\emph{hard} to \emph{soft} spectrum) and  $\eta({}^{252}Cf) / \eta({}^{88}Br)$ are presented. In addition, estimations of the efficiency for single ($\eta_{1n}$) and double ($\eta_{2n}$) \bdel neutron emission are also presented. For single emission, $\eta_{1n}$ corresponds to the efficiency averaged for ${}^{88}Br$ and ${}^{84}Ga$. For double emission, the efficiency is estimated from $\eta_{2n}=\eta^2_{1n}$. Inspection of table \ref{tab:TABnSpectra} shows that the methodology presented in this work has led to optimized configurations with a very flat response for both representative \bdel neutron spectra. Even for the harder spectra from a ${}^{252}Cf$ source, deviations less than 10\% are obtained for all the optimized configurations. When comparing results for the ratio ${\eta({}^{84}Ga)}/{\eta({}^{88}Br)}$, $\eta_{1n}$ and $\eta_{2n}$ in table \ref{tab:TABnSpectra}, a rather poor performace is observed for the flattest efficiency configuration with respect to the other ones. Therefore, flattest efficiency configuration is excluded from the further analysis for optimization of the detector in compact mode.

\section{Compact BRIKEN design}\label{sec:compact}

\begin{table}
\begin{center}
 \caption{Response of the optimized compact configurations to different neutron spectra.}
\label{tab:TABnSpectra_compact}
\setlength{\hcolw}{5pt}
\begin{tabular}{lccccc}
\noalign{\smallskip}\hline\noalign{\smallskip}
Configuration & \emph{F(5MeV)} & $\dfrac{\eta({}^{84}Ga)}{\eta({}^{88}Br)}$ & $\dfrac{\eta({}^{252}Cf)}{\eta({}^{88}Br)}$  & $ \eta_{1n}$ & $\eta_{2n}$\\ 
\noalign{\smallskip}\hline\noalign{\smallskip}
\emph{Good compromise flatness/efficiency}& 1.157 & $1.004 \pm 0.002$ & $0.957 \pm 0.002$  & $69.6\%$ & $48.4\%$\\ 
\emph{Maximum efficiency up to $1$ and $5MeV$}& 1.285 & $0.999\pm 0.002$ & $0.916 \pm 0.002$ &  $75.8\%$ & $57.5\%$\\ 
\noalign{\smallskip}\hline\noalign{\smallskip}
\end{tabular}
\end{center}
\end{table}

The transformation from hybrid to compact mode is achieved by removing both clover detectors and filling up the empty space with HDPE plugs. The $Z$-position of concentric tubes type E are modified by putting both ends in contact. As shown in fig. \ref{fig:G4detGeometry}.c, additional tubes of type N are placed in the middle region of the moderator. A further optimization of the configuration of these tubes has been carried out. For this purpose, the $XY$-positions of type N tubes   have been parameterized separately for configurations \emph{good compromise efficiency/flatness} and \emph{maximum efficiency}. These parameterizations introduce a new parameter ($g_r$) which defines the horizontal separation between the tubes on the $XY$-plane. Along the $Z-axis$, the active volumes of type N tubes are centered with respect to the center of the moderator. Eighteen tubes have been used for the new parameterization. Thus, a total of 166 \Hetub tubes are used for the compact mode. 

The optimization procedure is similar to that used for the hybrid mode. Therefore, two optimized configurations have been finally selected for the compact mode. The neutron responses of these optimized configurations are presented in figs.   
\ref{fig:NeutronSensitivityOptConf}.d and \ref{fig:NeutronSensitivityOptConf}.e. Examination of these results reveals that the transformation from the hybrid to compact mode yields an improvement in the flatness of the neutron response and the magnitude of the efficiency. In fact, for both configurations, the flatness value has a relative decrease of around 3\% and the neutron efficiency has an absolute increment of around 6\%.

The responses of both optimized configurations to different neutron spectra have been also studied. Results of these calculations are presented in table \ref{tab:TABnSpectra_compact}. The improvements of the flatness and efficiency for compact mode are reflected also when comparing tables \ref{tab:TABnSpectra} and \ref{tab:TABnSpectra_compact}. In fact, there is an increment of the ratio $\eta({}^{252}Cf) / \eta({}^{88}Br)$ towards the unity for the compact mode; in contrast, the ratio ${\eta({}^{84}Ga)}/{\eta({}^{88}Br)}$ remains constant for both modes.

\section{Final BRIKEN configuration}\label{sec:finalConf}
\begin{figure}[t]
\begin{center}
\includegraphics[width=0.5\textwidth,keepaspectratio]{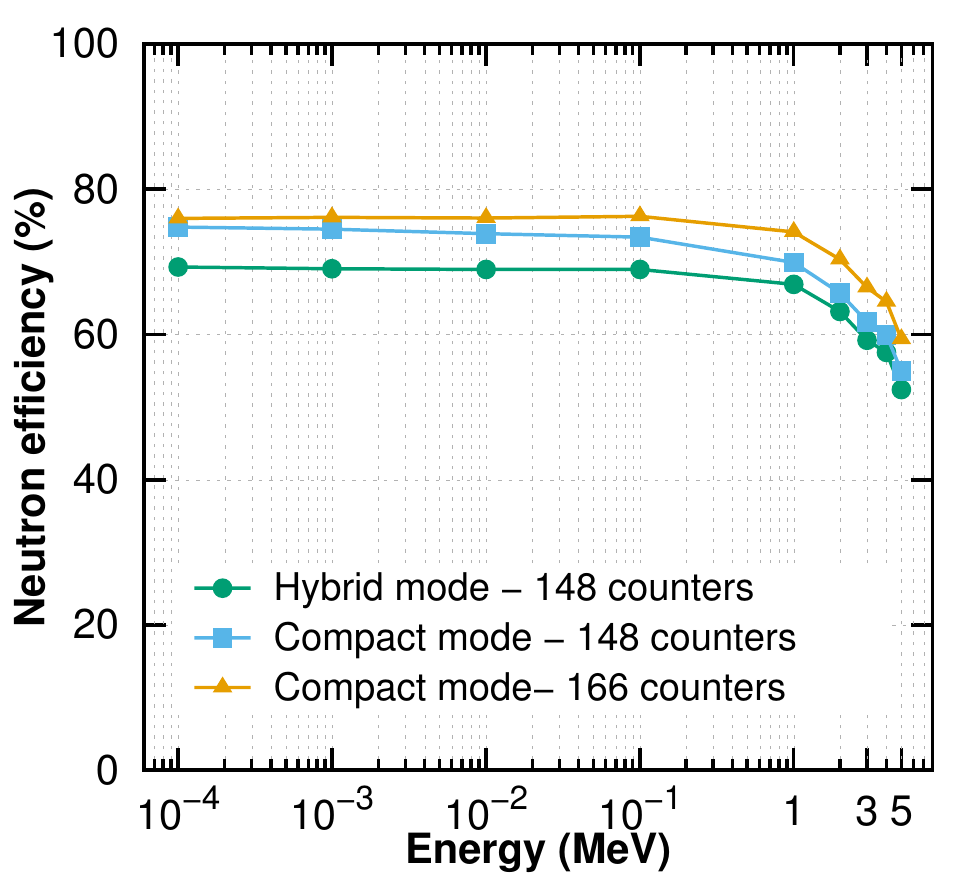}
\caption{\label{fig:DesignEff} Neutron response of the final configuration for operational modes hybrid (148 tubes) and compact (166 tubes). An additional compact configuration using 148 tubes is also shown.}
\end{center}
\end{figure}

Two optimized possible configurations have been proposed for the BRIKEN neutron detection system in hybrid mode. Both configurations can be transformed into the compact mode, which yields an overall increase of the efficiency and a slightly flatter neutron response. The performance of these two configurations, namelly \emph{good compromise flatness/efficiency} and \emph{maximum efficiency}, is summarized in fig. \ref{fig:NeutronSensitivityOptConf} and tables \ref{tab:TABnSpectra} and \ref{tab:TABnSpectra_compact}. \emph{Good compromise flatness/efficiency} configuration has a flatter response than \emph{maximum efficiency} configuration. However, a negligible impact is found when comparing the performances for the two representative \bdel neutron spectra (${}^{88}Br$ and ${}^{84}Ga$). The last is valid for both hybrid and compact mode. For the harder ${}^{252}Cf$ fission spectrum, the efficiency has  decrease less than $10\%$ relative for both configurations. On the other hand, when comparing the estimated values for $\eta_{1n}$ and $\eta_{2n}$, a clear advantage of the \emph{maximum efficiency} configuration for the compact mode is shown from the possibility to obtain an efficiency close to $58\%$ for double neutron emission.

The BRIKEN collaboration has finally chosen the \emph{maximum efficiency} configuration as the final decision for the BRIKEN neutron detector. This decision has been made based upon the physics programm, the above mentioned arguments and the technical advantages of each configuration. To summarize, the neutron response of the final configuration for both operational modes are presented in fig. \ref{fig:DesignEff}. An optional configuration for compact mode using only the 148 counters is also presented. This configuration is the compact one just obtained without using tubes type N. As can be seen from fig. \ref{fig:DesignEff}, by the action of transforming the setup into the compact mode with 148 tubes, the efficiency is increased in around 4.5\% absolute from lower energies up to $1~MeV$, but the flatness is impaired. The addition and optimization of tubes type N increase the efficiency up to 5~MeV and helps to recover the flat response.

\section{Discussion}\label{sec:discussion}
\begin{table}
\setlength{\tabcolsep}{3pt}
\begin{center}
 \caption{Comparison of different \bdel neutron detectors reported during the last fifteen years.}
\label{tab:detectorComparison}
\begin{tabular}{lcccccc}
\noalign{\smallskip}\hline\noalign{\smallskip}
Detector& \multirow{2}{50pt}{\centering Number of counters} &  \multirow{2}{40pt}{\centering Setup type}& $F(1MeV)$& $\big \langle \eta \big \rangle (1MeV)$ &$F(5MeV)$ &$\big \langle \eta \big \rangle (5MeV)$\\
& & & & & &\\
\noalign{\smallskip}\hline\noalign{\smallskip}
\multirow{2}{*}{\bf BRIKEN}& 166 & Compact & 1.029 & 75.7\% & 1.285 &71\% \\
			   & 148 & Compact & 1.07  & 73.3\% & 1.362 &67.6\% \\
			   & 148 &Hybrid & 1.036 & 68.6\% & 1.322 & 64.0\% \\
TETRA\cite{testovetal2016_3He} & 80 & Hybrid & 1.133 & 61.1\% & 1.842 & 51.6\% \\
MAINZ\cite{mathieuetal2012_new}${}^{**}$& 64 & Compact & 1.130 & 47.0\% & 2.245 & 39.9\% \\
NERO\cite{pereiraetal2010_neutron} & 60 & Compact & 1.157 & 43.1\% & 1.926 & 38.3\% \\
Hybrid-3Hen\cite{grzywaczetal2014_hybrid} & 48 & Hybrid & 1.103 & 36.8\% & 1.781 & 32.5\% \\
BELEN-20\cite{agramuntetal2015_characterization}& 20 & Compact & 1.062 & 46.2\% & 1.620 & 41.1\% \\
LOENI\cite{mathieuetal2012_new}${}^{**}$ & 18 & Hybrid & 1.016 & 17.4\% & 1.645 & 16.1\% \\
\noalign{\smallskip}\hline\noalign{\smallskip}
\multicolumn{7}{l}{\footnotesize ${}^{**}$Neutron response extrapolated for energies higher than 2MeV.}\\
\end{tabular}
\end{center}
\end{table}

A comparison of the BRIKEN detector performance with other \bdel neutron counters is presented in table \ref{tab:detectorComparison}. For each detector, neutron response functions from MC calculations have been reported in the corresponding publications. These functions are used for calculation of data presented in table \ref{tab:detectorComparison}. It is worth to mention that the table summarizes most of the setups for \bdel neutron measurements reported during the last fifteen years. 

Each detector in table \ref{tab:detectorComparison} has been designed addressing different technical requirement for the geometric setup, efficiency and flatness. Nevertheless, a pronounced enhancement of the efficiency with respect to the number of counters is observed for the BRIKEN neutron detector. The only exception to this trend is BELEN-20 detector, which setup has been particularly optimized to enhance the neutron detection efficiency \cite{agramuntetal2015_characterization}. Additionally, comparison of BRIKEN compact-148 and compact-166 configurations indicate that no important increment of the neutron efficiency can be achieved by include more counters to the configuration.

According to the corresponding publications, most of other detectors have been designed to have a nearly flat response up to 1~MeV. This is also reflected by the values of $F(1~MeV)$ in table \ref{tab:detectorComparison}. The flattest response for this energy range in the table corresponds to LOENI detector, which has been specially designed to have an energy independent efficiency \cite{mathieuetal2012_new}. In the second place appears the BRIKEN detector with a  $F(1~MeV)$ value similar to LOENI, but an  average efficiency up to 1~MeV fourfold higher. This fact highlights the impact of using a high amount of neutron detectors and the optimization methodology presented in this work.

One of the most interesting features of the BRIKEN neutron detector arises from examination of the flatness up to 5~MeV. BRIKEN compact-166 and hybrid-148 exhibit the flattest response in this energy range in comparison with other detectors in table \ref{tab:detectorComparison}. The relative difference in $F(5~MeV)$ values is significant, starting from around $25\%$ for LOENI and BELEN-20, up to $72\%$  in the case of the Mainz long-counter. In spite of that, the average efficiency up to 5~MeV of the BRIKEN detector is still the highest. Thus, the BRIKEN neutron detector is found to be the best compromise between high efficiency and flat response on table \ref{tab:detectorComparison}. To summarize, the high neutron efficiency, flat response up to 5~MeV, gamma-ray capabilities and flexibility of the set-up, make the BRIKEN neutron detector the state-of-the-art in neutron detectors for \bdel neutrons.

\section{Summary and final remarks} \label{sec:summary}

The conceptual design of the BRIKEN neutron counter is reported. This new detector has been conceived to measure beta-delayed neutron emission properties of a large amount of very neutron-rich nuclei at the RIKEN Nishina Center. 

In order to facilitate the design of the detector, a topological Monte Carlo optimization algorithm has been developed and implemented in \textsc{Geant4}. This methodology is based on the study of the transport and slowing down of neutrons in the HDPE moderator, the use of this information for parameterization of the counter array distribution, and the optimization of the array for high efficiency and flat response.

The design concept of the BRIKEN neutron detector is quite flexible and allows to operate the detector in hybrid and compact mode, while keeping a high efficiency and a nearly constant neutron response. In hybrid mode, the setup in composed by 148 \Hetub tubes and includes two clover detectors for high precision gamma spectroscopy ($\sim 1\%$ total photopeak efficiency at 1.33~MeV for each clover). In compact mode, the setup in composed by 166 \Hetub tubes and operates as a $4\pi$ neutron counter. The transformation from one mode to other is not difficult and can be achieved in between of experimental runs.

The response functions for both operational modes are very flat up to 1MeV and decrease slowly at higher energies. For example at 5~MeV, the relative decrease of the efficiency is less than 25\%. Moreover, for two representative spectra such as ${}^{88}Br$ beta-delayed neutron emitter ($\bar E_n=251 ~keV$) and the harder fission ${}^{252}Cf$ ($\bar E_n=2.28 ~MeV$), a relative decrease in the efficiency  of less than $10\%$ is found. For such a flat neutron response, the estimated efficiency for single \bdel neutrons in hybrid and compact mode are around 68.6\% and 75.8\%, respectively. Thus, efficiencies as high as 57.5\% can be obtained for double neutron emission. 

The BRIKEN neutron detector is joint effort to build a state-of-the-art detector for \bdel neutrons. This setup is going to be used for precise \bdel neutron measurements in very exotic nuclei. For this purpose, a collaboration involving 20 worldwide institutions has been stablished. The PE moderator has been builded during the first quarter of 2016 based on the design reported in this work. The full assembly and characterization of the BRIKEN detector will be done by the end of 2016. BRIKEN is an open project and will be available to the worldwide  nuclear physics community.


\end{document}